\documentstyle[12pt]{article}
\begin{document}

\def\a{\alpha}
\def\b{\beta}
\def\ch{\chi}
\def\d{\delta}
\def\e{\epsilon}
\def\f{\phi}
\def\g{\gamma}
\def\h{\eta}
\def\i{\iota}
\def\j{\psi}
\def\k{\kappa}
\def\l{\lambda}
\def\m{\mu}
\def\n{\nu}
\def\o{\omega}
\def\p{\pi}
\def\q{\theta}
\def\r{\rho}
\def\s{\sigma}
\def\t{\tau}
\def\u{\upsilon}
\def\x{\xi}
\def\z{\zeta}
\def\D{\Delta}
\def\F{\Phi}
\def\G{\Gamma}
\def\J{\Psi}
\def\L{\Lambda}
\def\O{\Omega}
\def\P{\Pi}
\def\S{\Sigma}
\def\U{\Upsilon}
\def\X{\Xi}
\def\T{\Theta}

\def\jb{\bar{\j}}
\def\yt{\bar{y}}
\def\wt{\bar{w}}
\def\zt{\bar{z}}
\def\vt{\bar{v}}
 \def\pp {\partial }
\def\pb {\bar{\partial }}
\def\be{\begin{equation}}
\def\ee{\end{equation}}
\def\ben{\begin{eqnarray}}
\def\np{\gamma^{\m }\partial_{\m } }
\def\een{\end{eqnarray}}
\def\gt{\tilde{g}}
\addtolength{\topmargin}{-0.8in}
\addtolength{\textheight}{1in}
\hsize=16.5truecm
\hoffset=-.5in
\baselineskip=7mm

\thispagestyle{empty}
\begin{flushright} \ April \ 1994 \\
KHTP-94-05 /SNUCTP 94-31 \\
\end{flushright}
\begin{center}
 {\large\bf  Spontaneous Breaking of Parity \\ in 2+1-Dimensional Thirring
Model
 }\\[.1in]
\vglue .5in
Y.M. Ahn, B.K.Chung, J.-M.Chung and Q-Han Park
\footnote{ E-mail address; qpark@nms.kyunghee.ac.kr }
\vglue .5in
{\it Research Institute for Basic Sciences \\
and \\
Department of Physics, Kyunghee University\\
Seoul 130-701, Korea}
\\[.2in]
{\bf ABSTRACT}\\[.2in]
\end{center}
\vglue .1in
A new aspect of the vacuum structure of 2+1-dimensional Thirring model is
presented. Using the Fierz identity, we split the current-current four-Fermi
interaction in terms of a matrix valued auxiliary scalar field and compute its
effective potential.  Energy  consideration shows that contrary to earlier
expectations, parity in general is  spontaneously broken at any finite order of
N,
 where N is the number of the two component spinors.  In the large N limit,
there
 does not exist a stable vacuum of the theory  thereby making the application
of
 the large N limit to Thirring model dangerous.
A detailed analysis for parity breaking solutions in N=2,3 cases is given.

\newpage

It is well known that 2+1-dimensional QED admits dynamical mass
generations.[1]-[3]
Solving the Dyson-Schwinger gap equation of the theory, Appelquist et al. [4]
have argued that in the large N limit, where N is the number of two-component
fermions,  masses are dynamically generated in such a way to preserve the
overall parity even though each individual mass term violates parity symmetry.
Recently, similar analysis has been applied to   2+1-dimensional Thirring
model of vector-vector type four-Fermi interactions with a same type of
dynamical
mass generations.[5] The four-Fermi interaction model  in 2+1 dimensions  are
also known to be renormalizable in the 1/N-expansion.[6]

In this Letter, we bring attention to the danger of this type of analysis,
particularly in the case of 2+1-dimensional Thirring model.  We analyze
symmetry-breaking patterns in 2+1-dimensional Thirring model for finite
N as well as for the large N limit. Using the Fierz identity,
we split the current-current four-Fermi interaction
in terms of a matrix valued auxiliary scalar field  and instead of solving
the  Dyson-Schwinger  gap equation, we compute the effective potential  of
the auxiliary field to study the dynamical mass generations.   Energy
consideration shows that contrary to earlier expectations, overall parity in
general is  spontaneously broken at any finite order of N.   In the large N
limit, there does not even exist a stable vacuum of the theory which  brings
doubts to  the 1/N-analysis of 2+1-dimensional Thirring model.
The discrepancy of our result with earlier works[4] based on
the Dyson-Schwinger equation arises because in  [4],
 the  overall parity-breaking solution was discarded in the large N limit
analysis since it exceeded the  ultraviolet  cut-off. However, our result shows
that
for finite N and for certain range of coupling constants the
parity-breaking solution becomes a true vacuum. In the large N limit,
it exceeds the ultraviolet cut-off as before which however makes other
solutions
semi-classically unstable. We argue that similar difficulty also arises for
2+1-dimensional QED.

Consider the 2+1-dimensional Thirring model given  by the Lagrangian
\be
L = \jb^{i}i\np \j^i - {g \over 2N}(\jb^i \g_{\m }\j^i )^{2 }
\ee
where $\j^{i}$ are two-component spinors and $i$ runs over from 1 to N.
In two-component representation, the Dirac $\g $ matrices are given in terms
of the Pauli matrices,
\be
\g^{0} = \s_{2} \ , \ \g^{1} = i\s_{3} \ , \ \g^{2} = i\s_{1} \ .
\ee
Note that the Lagrangian in Eq.(1) is invariant under the global $U(N)$
transformation as well as under the parity transformation $P$,
\ben
P:(x, y, t) &\rightarrow & (-x, y, t) \nonumber \\
\j^{'}(x^{'}) & \rightarrow & \j (x) = P\j (x) P^{-1} = \s_{1}\j (x) \ .
\een
It is believed that due to the strong  coupling of fermions, the parity
symmetry
breaks down spontaneously at the quantum level which generates dynamical masses
for fermions. However, it was argued that for  even  number of fermions $N=
2n$,
half of the fermions $\j^{i} \ ; i = 1,..., n$ get mass $m$ while the other
half
$\j^{i}  \ ; i = n+1, ... , 2n$ get mass $-m$ so as to preserve the overall
parity $P_{4} = PZ_{2}$ where $Z_{2}$ mixes fermions:[5]
\be
Z_{2}\j ^{i}(x) \rightarrow \j^{n+i}(x) \ , \ Z_{2}\j^{n+i}(x)
\rightarrow \j ^{i}(x) \  ; i = 1,...,n \ .
\ee
This argument is based on the Dyson-Schwinger equation method in the large N
limit which gives rise to the same type of  dynamical parity breaking for
2+1-dimensional QED.[4]
In the following, we show that contrary to the above argument the overall
parity $P_{4}$ is in fact broken. To do so, we first note that the action
Eq.(1)
can be written via Fierz transformation
\be
L = \jb^{i}i\np \j^i + {g \over 2N}(\jb^i \j^i )^{2 } +{g \over N}\jb^{i}
\j^{j}
\jb^{j}\j^{i}
\ee
where the four-Fermi interactions can be splitted by introducing a
matrix-valued
auxiliary field $M_{ij}$
\be
L = \jb (i\np + M)\j  - {N \over 4g}trM^2  +{N \over 4g(N+2)}(tr M)^2 \ .
\ee
This reduces to the original Lagrangian when $M_{ij}$ is eliminated by
integration which  identifies $M_{ij}$ with $M_{ij} = {g \over
N}\jb^{k}\j^{k}\d^{ij}+
{2g \over N}\jb^{j}\j^{i}$.
In order to understand the struture of vacuum, we compute the effective
potential
for $M_{ij}$ which amounts to summing the one loop fermion diagrams at zero
momentum. Since the matrix $M$ is hermitian, $M$ may be diagonlized with real
components $M=diag(\l_1 , \l_2, ... ,\l_{N})$ and  the effective potential,
 invariant under the diagonalization, is given by
\be
V_{eff} = -{\L^{3} \over 6\pi^{2}}\sum_{i=1}^{N} ln  (1+{\l_{i}^2  \over
\L^{2}})
-{1 \over 3\pi^{2}}\sum_{i=1}^{N}|\l_{i}|^{3}(tan^{-1}{|\l_{i}|\over \L}
-{\pi \over 2}) + ({N \over 4 g}-{\L \over 3\pi^{2}})\sum_{i=1}^{N}\l_{i}^{2}
- {N\over 4 g(N+2)}(\sum_{i=1}^{N}\l_{i})^2
\ee
where a cut-off has been introduced, after Wick rotation, at $k^2 = \L^2$ in
order that the integral be well defined. In terms of the dimensionless quantity
$x_{i} \equiv \l_{i}/\L $ and the rescaled coupling constant
$\gt \equiv  2\L g/3\pi^2 $,
 Eq.(7) becomes
 \be
\tilde{V}_{eff} = -\sum_{i=1}^{N} ln  (1+x_{i}^2 )
-2\sum_{i=1}^{N}|x_{i}|^{3}(tan^{-1} |x_{i}|
-{\pi \over 2}) + ({N \over \gt }- 2 )\sum_{i=1}^{N}x_{i}^{2}
- {N\over \gt (N+2)}(\sum_{i=1}^{N}x_{i})^2 \ .
\ee
In the following, we analyze various symmetry-breaking patterns of the
potential $\tilde{V}_{eff}$
according to the value of coupling constant $\gt $. First consider the case
$N=2$. The classical stability of the perturbative vacuum $(x_{1} = 0,
x_{2}=0)$
 is governed by the second derivatives of the potential $\tilde{V}_{eff}$ at
 $(0,0)$:
 \be
 A \equiv {\pp^{2}\tilde{V} \over \pp x_{1}^2 }|_{(0,0)} = -6+{3 \over \gt } \
, \
B \equiv {\pp^{2}\tilde{V} \over \pp x_{1}\pp x_{2} }|_{(0,0)} = -{1 \over \gt
}
\ , \ C \equiv {\pp^{2}\tilde{V} \over \pp x_{2}^2 }|_{(0,0)} = -6+{3 \over \gt
}
\ee
which imply that the perturbative vacuum is a local maximum if $B^2 - AC < 0$
and $A+C <0$, a local minimum if $B^2 -AC < 0$ and $A+C >0$, or a saddle
point if $B^2 -AC >0$. Thus, the potential $\tilde{V}_{eff}(x_{1},x_{2})$ and
the perturbative vacuum $(0,0)$ have the following properties;

i) $\gt < 0$:  $\tilde{V}_{eff}$ is bell-shaped  and $(0,0)$ is the absolute
maximum of  $\tilde{V}_{eff}$ so that there is no stable vacuum.

ii) $0 < \gt < 1/3$:  $\tilde{V}_{eff}$ is cup-shaped and $(0,0)$ is the
absolute
minimum of  $\tilde{V}_{eff}$ so that $(0,0)$ is a stable vacuum.

iii) $1/3 < \gt < 2/3 $:  There are  two local minima at $(m,m)$
 and $(-m,-m)$ for $0<m<1$.  $\tilde{V}_{eff}$  is  double-well shaped when
restricted
on the line $x_{1} = x_{2}$  and U-shaped when restricted on the line $x_{1}
= - x_{2}$ and $(0,0)$ is a saddle point of  $\tilde{V}_{eff}$. Thus,
$(\pm m,\pm m)$ become stable vacua which break parity symmetry spontaneously.

iv) $2/3 < \gt $: There are four local minima at $(m_{1},m_{1}), (-m_{1},
-m_{1})$
and $(m_{2}, -m_{2}), (-m_{2},m_{2})$.  $\tilde{V}_{eff}$ is double-well shaped
when restricted  on the line $x_{1}=x_{2}$ as well as on the line
$x_{1}=-x_{2}$.
$m_{1} > m_{2}$ and  $\tilde{V}_{eff}(\pm m_{1},\pm m_{1}) <
\tilde{V}_{eff}(\pm m_{2}, \mp m_{2})$. Thus $(\pm m_{1}, \pm m_{1})$
become stable vacua which break parity symmetry spontaneously.
However numerical computation shows that for $\gt \buildrel{\displaystyle > }
\over {\displaystyle_\sim } 1.5$, $m_{1} >1$.
This implies that the dynamically generated
mass exceeds the cut-off $\L$ and therefore $(\pm m_{1}, \pm m_{1})$ are not
sensible vacua. Nevertheless, this makes $(\pm m_{2},\mp m_{2})$
semi-classically
unstable so that there is no stable vacuum in this case.

For $N=3$, similar analysis leads to the following properites of the potential;

i) $\gt < 0$: $(0,0,0)$ is the absolute maximum of  $\tilde{V}_{eff}$ so that
there is no stable vacuum.

ii) $ 0< \gt < 2/5 $: $(0,0,0)$ is the absolute
minimum of  $\tilde{V}_{eff}$ so that $(0,0,0)$ is a stable vacuum.

iii) $2/ 5 < \gt < 1$: $(0,0,0)$ is a saddle point and there are two local
minima at $(\pm m,\pm m ,\pm m)$ which  become stable vacua  breaking parity
 spontaneously.

iv) $1 <\gt $: $(0,0,0)$ is a local maximum and there are six local minima with
same potential value and two absolute minima at $(\pm m , \pm m , \pm m)$
breaking parity spontaneously which become sensible vacua for
$\gt \buildrel{\displaystyle < }
\over {\displaystyle_\sim }  1.9$.

In order to understand the large N  behavior, we first consider the even N case
and then take the large N limit.  For $N=2n$, we look at the special sector of
the domain of the potential: $x_{1}=x_{2}= \cdots =x_{n} =x$ and
$x_{n+1}=x_{n+2}=\cdots
=x_{2n}=y$ where the symmetry breaking vacuua are expected to arise. Then the
effective potential becomes
\ben
{2\over N} \tilde{V}_{eff}(x,y) &=& -[ ln  (1+x^2 ) + ln (1+y^2 )]
-2[|x|^3 (tan^{-1} |x| -{\pi \over 2}) + |y|^3 (tan^{-1} |y|
-{\pi \over 2})] \nonumber \\
&& + ({N \over \gt }- 2 )(x^{2} + y^{2})
- {N^2 \over 2\gt (N+2)}(x+y)^2 \ .
\een
In terms of a rescaled coupling constant $g^{'} \equiv 2\gt / N $, the
structure of the effective potential ${2\over N} \tilde{V}_{eff}(x,y)$ is
the same as that of the $N=2$ case with the replacement of range of
$g^{'}$: i) $g^{'}< 0 \ \ ,$
ii) $0 < g^{'} < 4 / 3(2+N)  \ \ ,
$ iii) $ 4 / 3(2+N)< g^{'} < 2/ 3 \ \ , $
 iv) $2 / 3 < g^{'}$. For large N, the potential
 ${2\over N} \tilde{V}_{eff}$ possesses local minima along the line $x=-y$ when
$g^{'} > 2/ 3 $. However,  in the limit $N \rightarrow \infty$ the potential
becomes unbounded below when
restricted on the line $x=y$ for all values of $g^{'}$. Therefore, there does
not exist a stable vacuum  in the large N limit! At first sight,  this result
seems to be in  contradiction to earlier works based on the Dyson-Schwinger
equation  method.  In the Dyson-Schwinger type analysis with  the
1/N-approximation,   local minima  arise along the line $x = -y$ which
agrees with our result.  However  another set of solutions to the
Dyson-Schwinger equation,  which arise on the line
$x=y$ and also in  agreement  with our results, were discarded because
they exceed the momentum cut-off range.  Our results show that these
solutions can not be  simply discarded since they make other local
minima semi-classically unstable.
Finally,  we note that even though our analysis is only for the case of  2+1
dimensional   Thirring model, similar vacuum instability in the large N limit
may exist in  other 2+1-dimensional model as well.  In particular the vacuum
 struture  of 2+1-dimensional QED  was essentially the same as that of the
Thirring model in the  Dyson-Schwinger approach which suggests that the
1/N-approximation of 2+1-dimensional QED is as well dangerous.
  \vglue 1.in
{\bf    ACKNOWLEDGEMENT }
\vglue .3in
We are greatly indebted to Dr. S.H.Park for discussion and explanation of
their work.  We also thank Profs. H.J.Shin and S.Nam for their help.
This work was supported in part by the program of Basic Science Research,
Ministry of Education,  and by Korea Science and Engineering Foundation.

\vglue .2in

\def\Item{\par\hang\textindent}

{\bf REFERENCES }
\vglue .1in
\Item {[1]} W.Siegel,Nucl.Phys.{\bf B156}, 135 (1979); J.Schonfeld,
{\it ibid.} {\bf B185}, 157 (1981).
\Item{[2]} R.Jackiw and S.Templeton, Phys.Rev.{\bf D 23}, 2291 (1981).
\Item{[3]} S.Deser, R.Jackiw and S.Templeton, Ann.Phys.(N.Y.)
{\bf 140}, 372 (1982).
\Item {[4]} T.Appelquist, M.Bowick, D.Karabali, and L.Wijewardhana,
Phys.Rev.{\bf D33}, 3704 (1986); R.Pisarski, {\it ibid.} {\bf D29}, 2423
(1984); T.Appelquist, M.Bowick, D.Karabali, and L.Wijewardhana, {\it ibid.}
{\bf D33} 3774 (1986) .
\Item {[5]} D.K.Hong and S.H.Park, ``Large N analysis of 2+1 dimensional
Thirring model", SNUCTP 93-47
\Item {[6]} B.Rosenstein, B.J.Warr, and S.H.Park, Phys.Rev.Lett. {\bf 62}, 1433
 (1989); Phys.Rep.{\bf 205},59 (1991) and references therein.

\end{document}